\documentclass[]{spie}  
\pdfoutput=1

\newcommand{\unit}[1]{\ensuremath{\, \mathrm{#1}}}
 \usepackage{siunitx}
\usepackage{amsmath,amsfonts,amssymb}
\usepackage{graphicx}
\usepackage[colorlinks=false, allcolors=blue]{hyperref}

\title{Radiation effects on the Gaia CCDs after 30 months at L2}

\author[a,b]{Cian  Crowley}
\author[a,c]{Asier Abreu}
\author[a]{Ralf Kohley}
\author[d]{Thibaut Prod'homme}
\author[d]{Thierry Beaufort}
\affil[a]{European Space Agency, ESAC, P.O. Box 78, Villanueva de la Ca\~nada, 28691 Madrid, Spain}
\affil[b]{HE Space Operations for ESA}
\affil[c]{Elecnor Deimos Space for ESA}
\affil[d]{European Space Agency, ESTEC, Keplerlaan 1, 2201 AG,  Noordwijk, The Netherlands}

\authorinfo{Further author information: \\E-mail: cian.crowley@esa.int}

\begin{document} 
\maketitle

\begin{abstract}

Since the launch of ESA's Gaia satellite in December 2013, the 106 large-format scientific CCDs onboard have been operating  at L2. Due to a combination of the high-precision measurement requirements of the mission and the predicted proton environment at L2, the effect of non-ionizing radiation damage on the detectors was early identified pre-launch as potentially imposing a major limitation on the scientific value of the data. In this paper we compare pre-flight radiation-induced Charge Transfer Inefficiency (CTI) predictions against in-flight measurements, focusing especially on charge injection diagnostics, as well as  correlating these CTI diagnostic results with solar proton event data. We show  that L2-directed solar activity has been relatively low since launch, and radiation damage (so far) is less than originally expected. Despite this, there are clear cases of correlation between earth-directed solar coronal mass ejection events and abrupt changes in CTI diagnostics over time. These sudden jumps are lying on top of a rather constant increase in CTI which we show is primarily due to the continuous bombardment of the devices by high-energy Galactic Cosmic Rays. We examine the possible reasons for the lower than expected levels of CTI as well as examining the effect of controlled payload heating events on the CTI diagnostics. Radiation-induced CTI in the CCD serial registers and effects of ionizing radiation  are  also correspondingly lower than expected, however these topics are not examined here in detail. 

\end{abstract}


\section{INTRODUCTION}
\label{sec:intro}  

Gaia  is an ESA cornerstone (medium class) astrometry mission that was launched from Kourou in French Guiana in December 2013. Since the switch-on of the focal plane array (FPA) in early January 2014, the 106 large-format custom-made CCDs onboard have been operating in the  L2 radiation environment. Nominal operations were entered in July 2014 and the satellite will continue to operate in an orbit around L2 for at least the duration of the nominal mission lifetime of 5 years in addition to the 6 months of commissioning. Gaia rotates with a period of around six hours and precesses with a scan-law that allows it to slowly scan the entire sky with its two optical telescopes. The detection is carried out autonomously onboard and windows around the detected objects are readout from the CCDs and subsequently downlinked to ground (see Ref.~\citenum{2015AA...576A..74D} for further detail). Over the course of the nominal mission duration Gaia will  repeatedly observe more than one thousand million stars. A subset of results based on data acquired over the first year of nominal operations is currently being prepared and will be published in the form of a first intermediate data release to be published in late Summer 2016. In addition to positions and parallaxes, the final data release will also include colors, low-resolution spectra, photometry  and astrophysical parameters of every star, along with radial velocities and medium-resolution spectra of a subset of brighter objects. For details on the mission concept and science case see Ref.~\citenum{2001A&A...369..339P}. For post-launch astrometric performance predictions see Ref.~\citenum{2014EAS....67...23D} and for the latest  updates on the  data releases and the mission in general see \url{http://www.cosmos.esa.int/web/gaia}.

 Due to a combination of the high-precision measurement requirements of the mission and the pre-flight predicted proton environment at L2, the effect of non-ionizing radiation damage on the detectors was  identified pre-launch as potentially imposing a major limitation on the scientific value of the data. This damage to the  silicon lattice structure results in the generation of  charge  trapping sites  which cause a degradation in the efficiency in the transfer of charge packets from one pixel to the next (see, for example, Refs.~\citenum{2001sccd.book.....J},~\citenum{2001ITNS...48.1790H},~\citenum{2012MNRAS.419.2995P},~\citenum{2012MNRAS.422.2786H}). We refer to this as the Charge Transfer Inefficiency (CTI). CTI  causes alterations to the image shapes during readout from the CCD, which has the potential to cause significant biases in the measurement of stellar transit times and flux determinations if not modeled to sufficient accuracy. At this early point in the iterative data-processing cycles no CTI modeling has been included in the software pipeline, but in later cycles a fast forward model will be used during  the image processing, see Ref.~\citenum{2013MNRAS.430.3078S} for further details.

 The detector performances after two years at L2 (including a broad overview of the effects of the radiation environment) are described in Ref.~\citenum{detector_dr1}\footnote{This paper is part of the documentation that will accompany the first intermediate data release. This set of papers  will also contain the most up-to-date details on the status of mission, the data-processing and the scientific yield.}. In  this paper we provide a comparison of the results obtained from the in-flight diagnostics (up until launch + 30 months) to the pre-flight expectations (Sect.~\ref{sec:monitoring}). In Sect.~\ref{sec:factors} we   investigate and discuss the origin for the mis-match between the two. In Sect.~\ref{sec:heating_events} we analyze the observed changes in the CTI-diagnostics during controlled payload heating events. Finally, in Sect.~\ref{sec:discussion} we discuss  the implications of these studies and areas for further investigations and monitoring. However, before this, we present a brief overview of the Gaia CCDs and FPA in the following section.

\subsection{CCDs and the focal plane\label{sect:fpa}}

The Gaia FPA consists of 106 large area CCDs mounted on a Silicon-Carbide support structure, operated at a nominal temperature of $163$~K. The  devices  continuously operate in Time Delay \& Integration (TDI) mode where the parallel pixel clocking rate is  in synchronization with the satellite spin-rate.  Each CCD was custom-made by e2v technologies for the Gaia project and consists of $4500 \times 1966$ (parallel $\times$ serial) pixels of physical dimension $10 \times$~\SI{30}{\micro\metre}, which translates into~$\sim59 \times 177$~milli-arcseconds on the sky\footnote{In Gaia terminology the direction of transit is called the Along-Scan (AL) direction corresponding to what is usually called the parallel (or vertical) direction. We often refer to the serial (or horizontal) direction as the Across-Scan (AC) direction.}.
The TDI  clock rate is slightly less than~\SI{1}{\milli\second} (\SI{0.982}{\milli\second}), resulting in an effective exposure time of~\SI{4.42}{\second}  for a CCD transit. Of particular relevance in relation to hardware mitigation against radiation damage, the CCD design incorporates a structure that permits the injection of charge into the device to fill traps and a reset  the illumination history (i.e., it can be considered that all active traps are filled at the time of the charge injection). In addition, a special supplementary buried channel (SBC) is included in the pixel architecture. Its function is to confine small signal charge packets to a smaller transport volume than normal,  thus improving CTI at low signal levels, see Ref.~\citenum{2013MNRAS.430.3155S} for further detail. The total width of the scan sweep of the FPA (seven CCDs wide) corresponds to a scan width  on the sky of approximately 0.7$^{\circ}$.

$14$ of these devices are designated for onboard detection (in addition to   providing auxilary astrometric measurements), four for metrology (see Ref.~\citenum{2014SPIE.9143E..0XM}), $14$ for blue and red photometric measurements and $12$ CCDs are dedicated to spectral measurements around the CaII triplet in the near infrared. The remaining $62$ devices are part of what is known as Astrometric Field (AF) and are dedicated to low-noise astrometric measurements. It is these AF devices that we focus on in this paper. It can be noted that, although the pixel architecture and interfaces for all devices are identical, there are three different variants; a broadband device that is \SI{16}{\micro\metre} thick  (used for the AF); a variant optimised for blue wavelengths; and a thicker (\SI{40}{\micro\metre}) variant optimized for improved response at the red end. For a more detailed description of the CCDs and FPA see Refs.~\citenum{detector_dr1} and~\citenum{2012SPIE.8442E..1PK}.

\section{MONITORING RADIATION DAMAGE OF THE DETECTORS}
\label{sec:monitoring}

As well as serving to mitigate CTI by periodically filling trapping sites, the regular injection of blocks of charge into the CCDs provides a useful means of tracking the evolution of the CTI via analysis of both the First Pixel Response (FPR) and charge trails behind the injected lines (see Refs.~\citenum{2001ITNS...48.1790H} and~\citenum{detector_dr1} for further information on these concepts). These charge injection data are typically acquired via one of two means:

\begin{enumerate}

\item Special calibration activities which are periodically run onboard  with a typical cadence of three or four months. These data are acquired in order to monitor the CTI in the serial registers of the devices, as well as the parallel CTI for those devices which do not have periodic charge injection implemented during nominal operation (the effects of ionizing radiation are also monitored through a special activity). We do not discuss the results obtained from these data further in this paper except to note that both radiation-induced serial CTI and flatband voltage shift effects are low and consistent with the low levels that are seen for parallel CTI (as described in later sections of the paper). For an overview of the activities and the results during the commissioning phase see Ref.~\citenum{2014SPIE.9154E..06K}, for a presentation of the results obtained over the course of the first two years of the mission, see  Ref.~\citenum{detector_dr1}.

\item The regular downlink of windows that almost continuously sample the  nominal periodic charge injection profiles (and trails). For the AF devices four lines of charge are injected every 2~s and we deal exclusively with these data for the rest of this paper when referencing flight data.

\end{enumerate}

\begin{figure}
	\begin{center}
	\includegraphics[height=6.4cm]{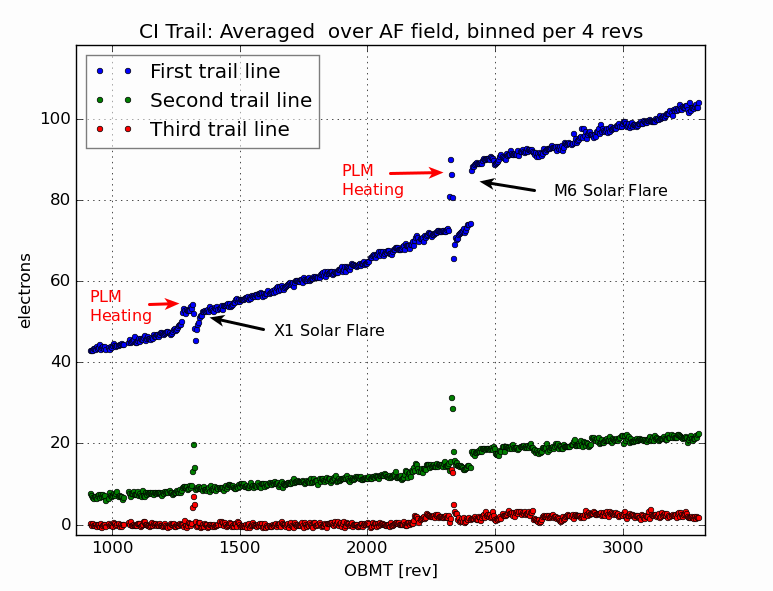}
	\includegraphics[height=6.4cm]{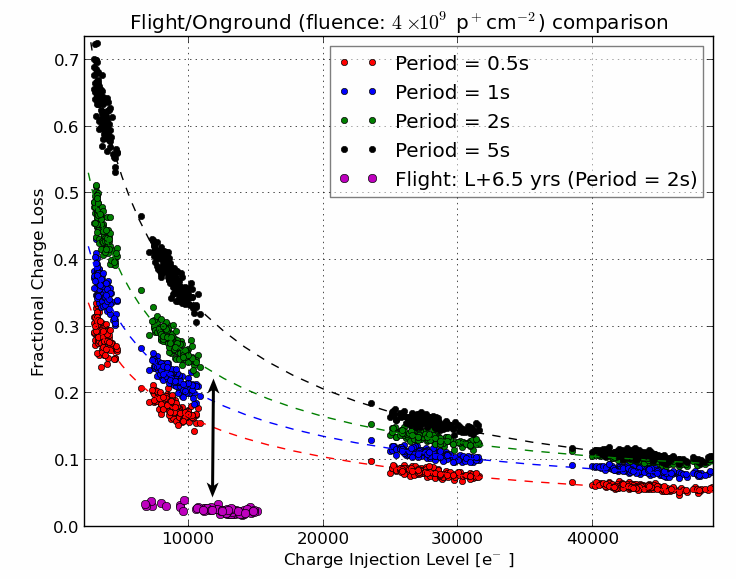}
	\caption{\label{fig:fpr}  {\bf{Left:}} Evolution of the median values over 24 hours for all AF CCDs of the first three trailing samples after charge injection. Monitoring the charge in  the trail behind the charge-injected rows permits the tracking of the parallel CTI over time. Clearly apparent on the plot are two distinct components contributing to the CTI increase, a rather linear increase and two step increases that can be traced to solar ejections. Note that the effects of two controlled decontamination events are also visible. The x-axis spans a little under two years.   {\bf{Right:}} A comparison of the fractional charge loss derived in-flight (after extrapolation to 6.5 after launch) to charge loss values measured on-ground using an irradiated device. It can be observed that the extrapolated in-flight measurements are approximately an order-of-magnitude less than the on-ground results. The comparison between both provides an estimate of the fluence deposited by Non-Ionizing Energy Loss (NIEL) processes during the mission so far, see text for further detail. 
}
	\end{center}
\end{figure}

Analysis of the charge injection data from the AF devices permits the continuous tracking of the status of the CTI level for all of these detectors. For example, in the left panel of  Fig.~\ref{fig:fpr} the number of electrons in the first, second and third pixel rows behind the charge injection block are plotted as a function of time into the mission (the units are represented in terms of OnBoard Mission Timeline (OBMT) spacecraft revolutions, or, roughly, six hours units). The datapoints show the median values for all pixel columns on all AF CCDs over a $24$ hour period. The rather constant degradation in CTI is readily visible, as well as the effects of two solar flare events. Also marked on the plot are the effects of two controlled heating events where the trailing increases dramatically and the shape of the release trail is temporarily modified (see Sect.~\ref{sec:heating_events} for a more detailed analysis). It should be noted that the sky-density also effects these values, higher sky-density results in more photo-electrons being produced in each device which results in more traps being kept filled between charge injections, resulting in small temporary dips in the trails/FPR values.

The level of the de-trapping trails (or, alternatively, FPR) also provides a platform for the comparison with similar diagnostics obtained from on-ground data on irradiated devices. 
There is an extensive on-ground dataset acquired using a dedicated Gaia detector test bench that is currently installed at the ESA/ESTEC
site in The Netherlands. 
The dataset consists of data obtained from Gaia flight-model and engineering-model CCDs which were irradiated to known proton fluences. These datasets can then be used to compare the pre-flight CTI predictions to results from in-flight measurements in order to estimate the NIEL fluence  accumulated so far at L2. 
Shown in the right panel of Fig.~\ref{fig:fpr} are the fractional charge loss values (simply the FPR values normalized by the injection level) averaged over the AF CCDs  after extrapolation to launch + 6.5 years (the nominal mission duration plus half a year commissioning as well as room for a possible mission extension of one year, these values were chosen in order to ease comparison to the pre-flight predictions which were carried out using the same mission duration). Also plotted  (against injection level) are the fractional charge loss values for a number of different charge injection periods that were obtained on-ground from an engineering model AF device that was irradiated at room temperature to a fluence of $4 \times 10^{9} ~\unit{ p^{+} / cm^2}$ of $10$~MeV NIEL equivalent protons. It can be observed that the extrapolated end-of-mission values are approximately an order of magnitude less than the values obtained from the on-ground dataset.

In the radiation sector analysis that was carried out prior to launch, it was found that the  $10$~MeV equivalent proton fluence that was predicted for the average of the AF after 6.5 years at L2 was $3.1 \times 10^{9} ~\unit{ p^{+} / cm^2}$. This analysis was carried out using the solar proton fluxes predicted by the model described in Ref.~\citenum{1993JGR....9813281F} (Galactic Cosmic Rays were neglected) at the  90\% confidence level. Therefore, by examining  the amount that we need scale down the on-ground fractional charge loss values  by in order to match the extrapolated $6.5$~year mission flight values, we conclude that we are on track to  accumulate $\sim 7 - 8 $~times less damage than was predicted by the sector analysis.

\section{FACTORS AFFECTING LOW OBSERVED LEVELS OF RADIATION DAMAGE}
\label{sec:factors}

Obviously the lower than expected CTI measurements are good news for the science goals of Gaia, but in order to gain an understanding of the origin of the discrepancy, in this section, we attempt to disentangle the different factors that may play a role.

\subsection{High straylight levels on Gaia}
\label{sec:straylight}

During the commissioning phase of Gaia it was discovered that there were unexpected and  significant levels of straylight affecting the focal plane. There were found to be two components, one related to diffraction of solar light around the sunshield where this component varies rather slowly over time. The second component originates from off-axis astrophysical sources (due to non-optimal baffling) and varies dramatically over the course of a six hour satellite revolution, and also from CCD-to-CCD. As described in Ref.~\citenum{2014EAS....67...23D}, the median level of straylight for all CCDs is $\sim5.5 \unit{e^{-} \text{pixel}^{-1}\text{s}^{-1}}$, with 90\% of the distribution less than  $20 \unit{e^{-} \text{pixel}^{-1}\text{s}^{-1}}$.   

Although these high levels  of straylight add photon noise to the measurements, they must certainly also be keeping some trapping sites filled, thus mitigating CTI. In fact, for some HST observations low levels of optical background are applied to the CCD on purpose (post-flash) before readout in order to mitigate the CTI (Ref.~\citenum{2015ASPC..495..323B}). However, the relevant question here is: do the high straylight levels onboard Gaia explain the discrepancy between pre-flight CTI predictions and the observed levels?

\begin{figure}
	\begin{center}
	\includegraphics[width=\columnwidth]{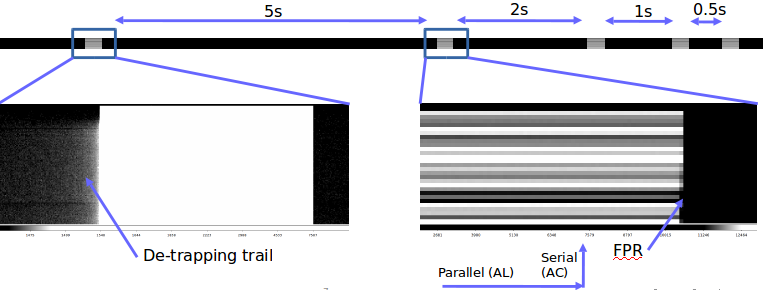}

	\caption{\label{fig:setup}  Example of a portion of the on-ground dataset that was acquired in order to investigate the effect of high straylight levels on the CTI diagnostics that are used onboard Gaia. Shown in the top panel is a section of one acquisition of 162 pixels over $16~000$~pixel rows, corresponding to $\sim 16$~s (it should be noted that we also refer to a row of pixels as a TDI line). The blocks of  pixel rows containing charge injections are clearly visible. The readout direction is from left to right (parallel) and from bottom to top (serial). Shown at bottom left is a zoom-in around on of the charge injection blocks where the de-trapping trail is visible using a logarithmic  stretch. At bottom right is a zoom-in around another charge injection block where the trapping from the first rows of injected charge is apparent (FPR). Many similar runs were acquired and the charge injection level and the optical background level were varied in order to explore the effect of the free parameters.
}
\end{center}
\end{figure}

In order to answer this question, a number of tests were carried out on the Gaia test bench on the same irradiated device that was used to compare the in-flight CTI values to the on-ground levels. The fractional charge loss was measured for a number of injection levels, injection periods and optical background levels in order to assess the level of mitigation induced by the straylight levels on the CTI diagnostics. The acquired dataset is briefly described in Fig.~\ref{fig:setup} and its caption.

\begin{figure}
	\begin{center}
	\includegraphics[height=6.cm]{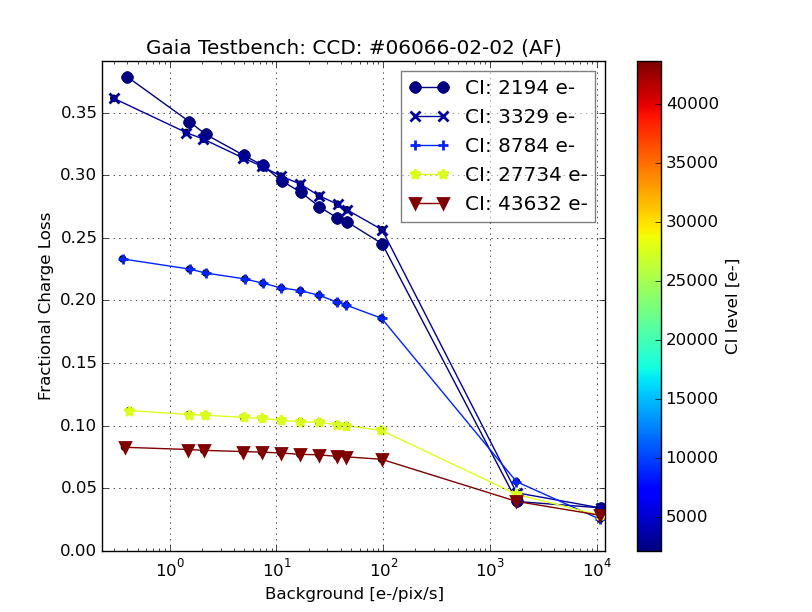}
	\includegraphics[height=6.cm]{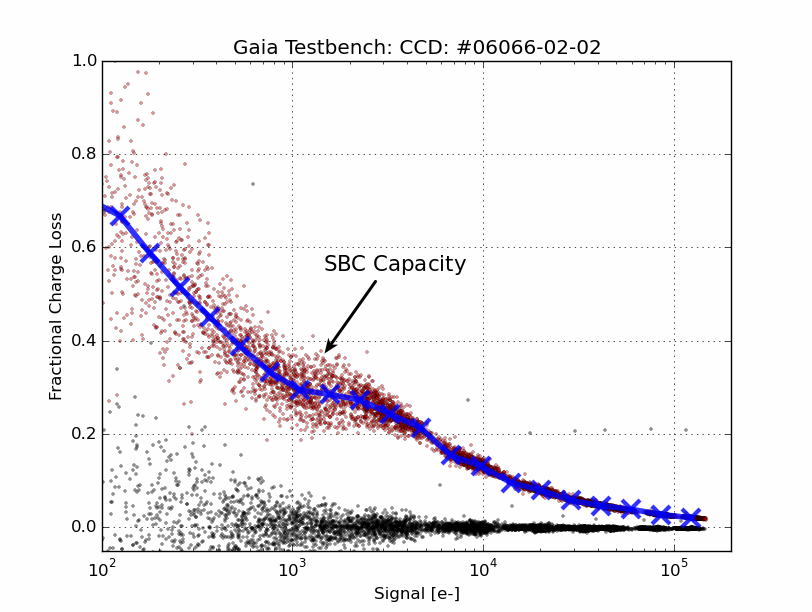}
	\caption{\label{fig:afstraylight} {\bf{Left:}} Results of on-ground experiments carried out in order to examine the effect of high straylight levels on CTI diagnostics as a function of charge injection level for an AF CCD. Each data-point is the result of averaging over many pixels ($127$) and acquisitions (nine). The fall-off of the measured FPR values with increased optical background is readily apparent. It can be noted that at extremely large levels of optical background ($\sim$~few thousand $\unit{e^{-} \text{pixel}^{-1}\text{s}^{-1}}$) the background level effectively keeps most traps permanently full and the level of the CTI diagnostic is drastically reduced. It can be noted that the curve for the lowest charge injection level actually crosses the curve for the next highest injection level at~$~10 \unit{e^{-} \text{pixel}^{-1}\text{s}^{-1}}$.  {\bf{Right:}} Fractional charge loss measurements for the AF device as a function of injection level. The results show a functional SBC with a capacity of $\sim1000$\unit{e^{-}}. The black datapoints correspond to acquisition of pixels from undamaged portions of the device.
}
	\end{center}
\end{figure}

The left panel of Fig.~\ref{fig:afstraylight} shows the fractional charge loss values measured on-ground (using a 2~s injection period in order to directly compare to the flight data from the AF CCDs) as a function of the level of optical background. The trends are plotted for a range of differing charge injection levels. The fall-off with fractional charge loss with increased background level is apparent for all injection levels, indeed it can be observed that it does not fall off very dramatically up to $\sim1000$\unit{e^{-}} (note the log scale of the x-axis). Indeed (from the right panel of Fig.~\ref{fig:fpr}), it can be observed that the average injection level across each AF CCD ranges from  $\sim 6~000 - 15~000 \unit{e^{-} \text{pixel}^{-1}}$ (there is also a large level of variation in the level from column-to-column within a CCD) and for this range of injection levels there is a relatively shallow drop-off in fractional charge loss with background level. Interestingly, for the lowest injection level it can be seen that the curve crosses that of the next highest injection level at a background level of $\sim 10 \unit{e^{-} \text{pixel}^{-1}\text{s}^{-1}}$ showing a different behavior in  the regime of the lowest level. This is likely related to the presence of a functioning SBC in this device with a capacity of $\sim 1000$\unit{e^{-}} (see the right panel of Fig.~\ref{fig:afstraylight} showing the change in fractional charge loss across the transition from the SBC regime into the buried channel regime for this device, see Ref.~\citenum{2013MNRAS.430.3155S} for further detail on the interpretation of these type of plots). Although the mean level of injection across the CCD pixels is over $2000 \unit {e^{-} \text{pixel}^{-1}}$ and therefore above the measured capacity of the SBC, due to the variation in the level from pixel-to-pixel (the variation is also higher at low injection levels), some of the pixels will be firmly in the SBC regime.

\begin{figure}
	\begin{center}
	\includegraphics[height=6.cm]{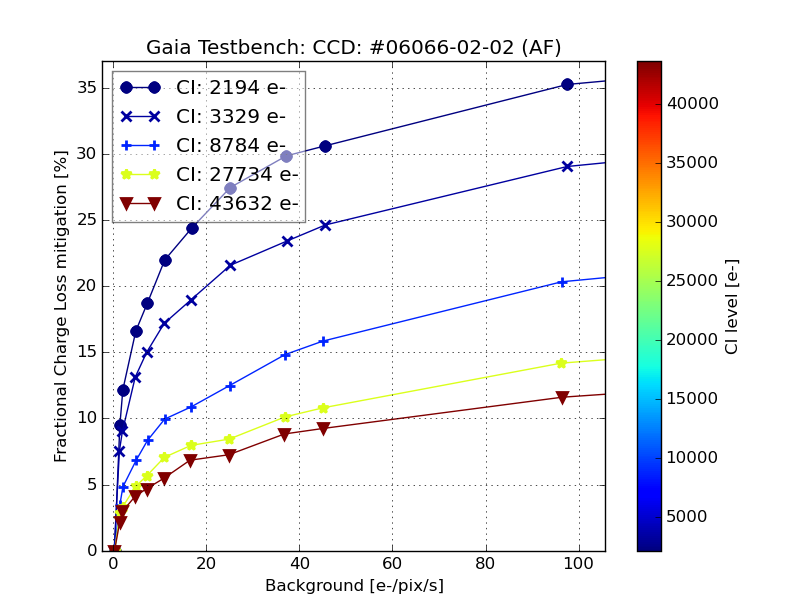}
	\includegraphics[height=6.cm]{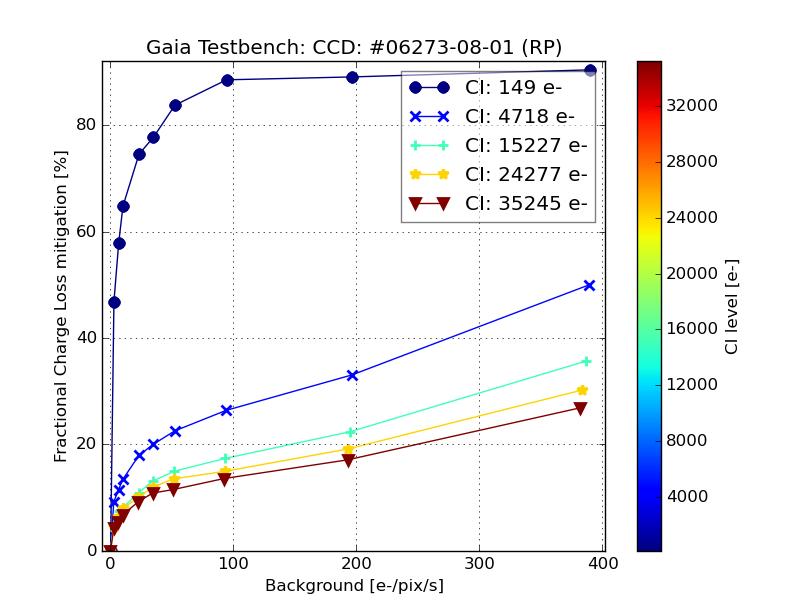}
	\caption{\label{fig:afrpmitigation} {\bf{Left:}} Percentage of mitigation on the zero-background fractional charge loss value as a function of optical background for the AF CCD tested on-ground.  {\bf{Right:}} Similar to the left panel, but for the RP device. After differing optical background levels and injection levels are taken into account, the results between both CCD variants are very similar.
}
	\end{center}
\end{figure}

In order to further quantify the effect of the optical background, displayed in the left panel of Fig.~\ref{fig:afrpmitigation} are the same data as displayed in the left panel of Fig.~\ref{fig:afstraylight}, but plotted in terms of the percentage of mitigation that the background level brings to the measured fractional charge loss relative to the case with no optical background. It is apparent that the addition of just a few $\unit{e^{-} \text{pixel}^{-1}\text{s}^{-1}}$ of background can rather significantly mitigate the measured fractional charge loss values, but the addition of increased background provides proportionally less mitigation. In fact similar results were found when analyzing on-ground   measurements of stellar image location biases pre-flight as a function of background level. As described in Ref.~\citenum{2011Obs...131..191B} it was found that a very  low level of background provided a reasonable level of mitigation, but that increased levels provided diminishing returns.

For the case of  flight-representative injection levels it is apparent that for background levels of $\sim 3 - 20 \unit {e^{-} \text{pixel}^{-1}\text{s}^{-1}}$  the mitigation is of the order of $\sim4 - 12$\%, which is significant, but does not come close to explaining the discrepancy between pre-flight and observed CTI levels.

For completeness, the right panel of Fig.~\ref{fig:afrpmitigation} shows similar results from on-ground tests, but carried out on a proton-damaged engineering model red-enhanced (thicker) device. Unsurprisingly, the results are similar to those for the AF device at comparable injection levels. Of some interest, however, is that the level of mitigation for the lowest injection level reaches almost 90\% at a background of $\sim100  \unit{e^{-} \text{pixel}^{-1}\text{s}^{-1}}$. The pixel-averaged injection level here was only $149 \unit {e^{-} \text{pixel}^{-1}}$, nominally firmly in the SBC regime for this device. This implies that the level of CTI-mitigation afforded by the large background levels may be actually  much greater for stellar images that lie in the SBC regime than for the actual charge injection profiles, which always contain too much signal in flight to lie in the SBC regime\footnote{It should also be noted that, due to the TDI integration from pixel-to-pixel operation of the CCDs, that even stellar images that are composed of electron packet sizes exceeding the SBC capacity at readout could well have undergone many  pixel transfers where the electron packets are below the SBC capacity.}. However, there is likely another process  to consider here in addition to the transport of electron packets through the SBC at low injection levels. It  is that at such low levels, the number of electrons that are injected into the device will likely not be  enough to keep the entire population of slow-release traps (where the emission time constant is much larger than the charge injection period) permanently filled, whereas this likely is the case at flight-representative injection levels. Therefore, at very low injection levels, the background is able to suppress many traps that are already suppressed by the charge injection at higher injection levels. Another implication of this is that, were it not for the periodic charge injection implemented onboard, the mitigation of CTI by straylight would be much higher, i.e., the straylight and periodic charge injections both heavily suppress the slow-release traps, Although, we note of course, that the straylight will also keep some of the faster-release traps suppressed, which will aid in the data-processing since these are the traps that  distort the image shapes the most. In summary, although the effect of high background levels  on the actual stellar images is yet to be  understood, it is demonstrated that the effect on the in-flight CTI diagnostic levels is modest.

\subsection{Operating temperature}
\label{sec:temperature}

CTI diagnostics are very sensitive to the operating temperature of the devices (indeed, this is quantified further in Sect.~\ref{sec:heating_events}). Therefore, if there were even a relatively small ($\sim$~few K) difference between the operating temperatures of the device on-ground and those in-flight, it could induce a large bias in the comparison of on-ground to flight NIEL estimates. Indeed, in-flight there is a relatively large temperature variation over the FPA of $\sim4.5$~K (see ~Ref.~\citenum{detector_dr1}). However, the average FPA temperature is $\sim164$~K (the CTI measurements are averaged from the whole AF of the FPA). This compares with a CCD operating temperature for the on-ground test of $163.5$~K. Therefore, the bias in the comparison due to temperature differences will be minimal (see Sect.~\ref{sec:fpr_heating} for more detail on the temperature dependencies).


\subsection{Irradiation temperature}
\label{sec:irr_temperature}

It is known (i.e., Ref.~\citenum{2005ITNS...52..519B}) that temperature of  irradiation (cryogenic versus room temperature, for example) and the thermal cycling of an irradiated device can change the trap populations and characteristics. Since the on-ground data were acquired using a device that was proton-irradiated at room temperature, it is conceivable that this could complicate (or even invalidate) the fluence comparisons between on-ground and flight measurements. However, a study is described in  Ref.~\citenum{2010ITNS...57.2035H} where a Gaia test device had portions irradiated at both room temperature and at $143$~K and the CTI diagnostics between the two were compared. It was found that the density of traps was approximately a factor of $2$ higher for the cold-irradiated case relative to the warm-irradiated case. However, when the authors break this down in terms of the characteristic emission times ($\tau _{e}$) of the traps, they find that the concentrations are $2-3$~times higher for the cold-irradiation case for traps with $\tau_e \lesssim 100$~ms. For those traps where $\tau_e \gtrsim 300$~ms the ratio is less than unity at $\sim0.75$. For the charge injection diagnostics being analyzed in this paper we are sensitive to traps that release within the charge injection period of 2~s. Therefore, we believe in this case that the factor~$2$ is probably an upper-limit for our purposes and that a warm-cold irradiation comparison for Gaia at these operating conditions is valid. In any case, even allowing for a factor of~$2$ difference, we note that the trap concentration should be higher for the cold-irradiated case (the opposite sense to what we observe), so this certainly can not explain the discrepancy between the expected versus predicted radiation damage.

\subsection{Comparison of radiation environment predictions to observations}
\label{sec:environment}

In this section we examine the predicted  radiation environment and compare it to what has been observed so far over the first $30$ months after launch. The pre-flight NIEL predictions  were carried out using the Interplanetary solar proton fluence model of Ref.~\citenum{1993JGR....9813281F}  which assumed a duration of solar maximum of $3.5$~years and the solar fluences were computed to the $90$\% confidence level (Galactic Cosmic Rays (GCRS) were neglected in the analysis due to their low flux levels). As previously mentioned, the resulting sector analysis  predicted an  average NIEL fluence value for the AF CCDs of $3.1 \times 10^{9} ~\unit{ p^{+} / cm^2}$ of $10$~MeV equivalent protons.

Of course, the use of a $90$\% confidence level will add a margin onto the most likely proton fluence level, that of the $50$\% confidence level. In order to assess the width of the this distribution we re-run the model, but for a simple shielding geometry (spherical geometry using 11~mm of aluminium shielding) for both the $90$\% and $50$\% confidence levels. After conversion of both fluence spectra to $10$~MeV NIEL equivalence it is found that the difference between both is a factor of $\sim3.8$. Therefore, if L2-directed solar proton activity was actually in the lower tenth percentile since launch then we would expect to see a factor of $\sim3.8 \times 3.8 \sim14$ times less CTI than was predicted, which is of the order of magnitude  which we observe. However, what of  the effect of the GCRs which were neglected in original analysis? Could they be the source of the near-linear component in the CTI degradation observed in the left panel of Fig.~\ref{fig:fpr} that has caused $\sim75$\% of  the observed CTI so far? 

Although the  continuous and isotropic flux of GCRs  varies a little over the course of a solar cycle (due to the changing extent of the protective interplanetary magnetic field and solar wind over the solar cycle), the flux can be considered constant to  first approximation when assessing NIEL fluences. Although the fluxes are low (a few particles per cm$^{2}~\rm{s}^{-1}$) we can assess the impact over a 6.5 year mission by comparing the end-of-mission GCR fluences to those predicted by the solar proton model.

\begin{figure}
	\begin{center}
	\includegraphics[height=6.cm]{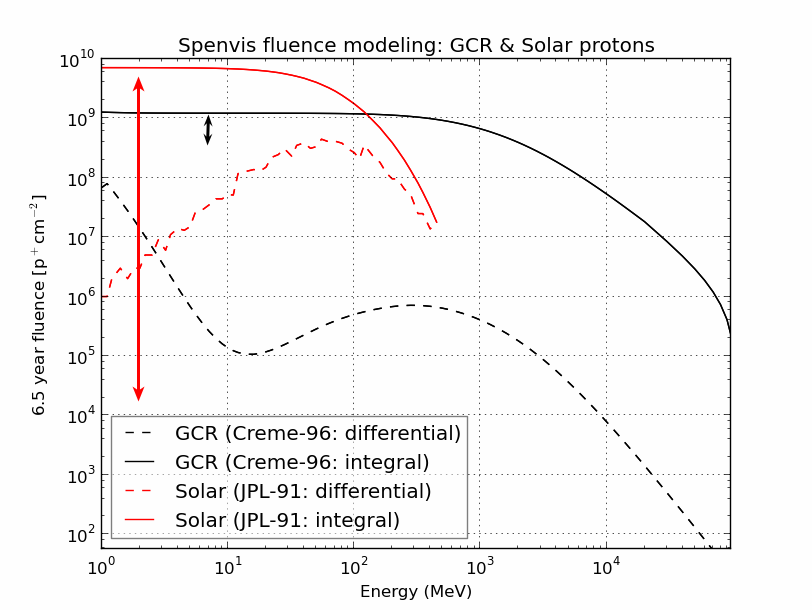}
	\includegraphics[height=6.cm]{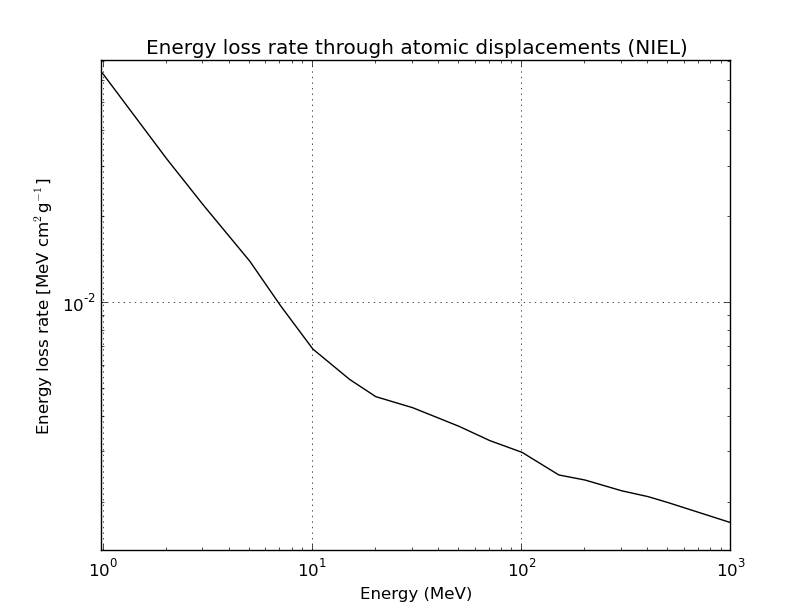}
	\caption{\label{fig:model} 
{\bf{Left:}} 
Proton environment models  produced via SPENVIS using the interplanetary solar proton fluence model of Ref.~\citenum{1993JGR....9813281F} (for solar protons, in red) and  Ref.~\citenum{1997ITNS...44.2150T} (for GCR protons at solar minimum, in black). The dashed curves show the differential fluences (in units of cm$^{-2}~\rm{s}^{-1}~\rm{MeV}^{-1}$) and the continuous curves correspond to the total fluences integrated over all higher energies.  The solar proton fluences are computed after shielding effects have been applied. No shielding has been applied to the GCR fluences. The arrows depicted on the plot signify the time variability of the fluxes of each component, where the solar fluxes vary dramatically over time and the GCR fluxes evolve slowly and by a small amount.
{\bf{Right:}}  
The energy loss rate through atomic displacements (NIEL) over a range of proton energies. The plot was generated using data  from Ref.~\citenum{1994ITNS...41.1974D}.
}
	\end{center}
\end{figure}

Displayed in the left panel of Fig.~\ref{fig:model} are the spectral energy fluence  distributions  (both integral and differential) after 6.5 years at  L2 of both the solar model ($90$\% confidence level, after shielding) and a GCR model (protons only, no shielding included). See the caption for further detail. This plot summarizes the expected proton environment, i.e., the GCR flux remains rather constant over time and peaks at high energies ($\sim500$~MeV). In contrast, the solar proton fluxes are highly variable over time, but are expected to average out (in this $90$\% confidence level model) to  provide an integrated fluence of  $\sim 7 \times10^{9}~\unit{ p^{+} / cm^2}$, which is $\sim7$ times higher than the integrated GCR fluence ($\sim 1 \times10^{9}~\unit{ p^{+} / cm^2}$), although peaking at a lower energy of $\sim50$~MeV. However, in order to compare the relative levels of displacement damage imparted by both sources it is required that the different energy peaks of each distribution be taken into account since the NIEL curve changes with proton energy, as shown in the right panel of Fig.~\ref{fig:model}.

An examination of the displayed NIEL curve shows that the energy loss at $50$~MeV will be about half  that at the commonly-used reference level of  $10$~MeV. In turn, the energy loss at $500$~MeV is about half again of what it is at $50$~MeV, so if we used these approximate values to normalise both the solar  GCR distributions to $10$~MeV NIEL equivalence, we derive fluence values of  $\sim \frac{7}{2} \times10^{9}~\unit{ p^{+} / cm^2}$ and $\sim \frac{1}{4} \times10^{9}~\unit{ p^{+} / cm^2}$ respectively. So even though the GCR flux rate is low, the displacement damage    is expected to be $\sim \frac{1}{14}$ times that of the solar proton events at the $90$\% confidence level. This is a non-negligible number, especially in the presence of low sun activity.

We can also examine how this number for the GCRs ($\sim 2.5\times10^{8}~\unit{ p^{+} / cm^2}$) compares to the end-of-mission-extrapolated fluence derived from flight data ($\sim75$\% of the pre-flight estimated AF average fluence divided by the observed factor of $\sim 7 - 8 $ found from the on-ground/flight comparison $\sim$$\frac{3.1}{7.5} \times 10^{9} ~\unit{ p^{+} / cm^2}$ $\sim$$  3 \times 10^{8} ~\unit{ p^{+} / cm^2}$ ). So it can be seen that the numbers match very well, and that the linear increase in CTI is well-explained by the displacement damage due to the GCRs.   It should be noted that the linear component of the CTI increase was analyzed for any evidence of a change in the slope over time, as might be expected to exist due to the reduction of the protective effect of the solar magnetic field as solar activity decreases. So far, there is no evidence of this effect in the data, however, a longer base-line is really needed in order to say something meaningful about this. In addition, it is shown in Ref.~\citenum{detector_dr1} that the differential shielding pattern across the CCDs in the AF is reflected in the magnitude of the step increases (solar events), whereas the slopes of  linear component does not reflect such a pattern. This confirms that the particles causing the linear increase are of higher energy than those causing the step increases.

However, since the GCRs are causing $\sim 75$\% of the parallel CTI, and these were neglected in the pre-flight predictions, then the CTI caused by solar proton events must be even lower than first assumed. Since they are only causing $\sim 25$\% of the damage accumulated so far, then the extrapolated end-of-mission NIEL fluence ($10$~MeV NIEL equivalence) for solar protons will be $\sim 25$\% of $\frac{3.1}{7.5} \times 10^{9} ~\unit{ p^{+} / cm^2}$ $\sim$$  1 \times 10^{8} ~\unit{ p^{+} / cm^2}$. This figure is $\sim30$ times less than that predicted pre-flight, and still eight times less than would have been predicted if  a $50$\% confidence level was used. So, has the  sun really been so inactive since the launch of  Gaia?

\begin{figure} [ht]
   \begin{center}
   \begin{tabular}{c} 
   \includegraphics[height=6.cm]{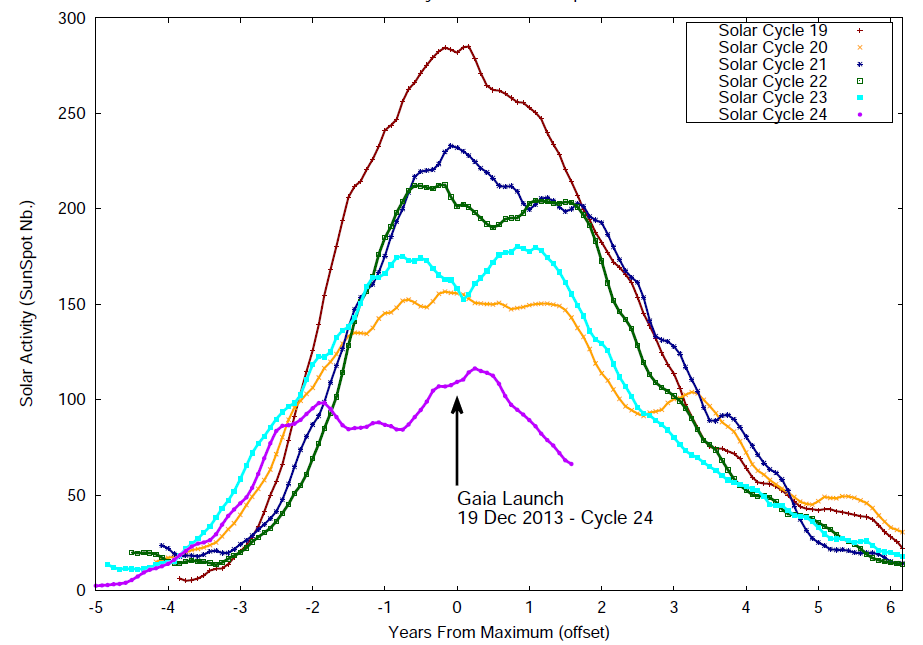}
   \includegraphics[height=6.49cm]{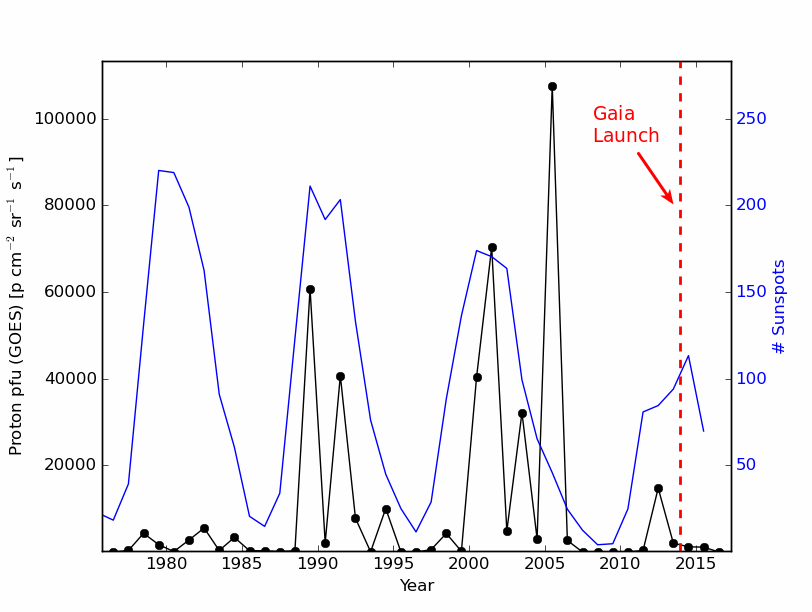}
   \end{tabular}
   \end{center}
   \caption{ \label{fig:solarActivity} 
{\bf{Left:}} Monthly-mean  sunspot counts (phase-wrapped and smoothed) for the last six solar cycles. The relative position of the Gaia launch is denoted for cycle $24$. Note the low level of sunspot activity during this cycle relative to the previous five (source data: WDC-SILSO, Royal Observatory of Belgium, Brussels). 
{\bf{Right:}} Yearly-averaged monthly  sunspot counts for the last four solar cycles are plotted (blue) along with yearly totals of the  proton fluxes directed towards earth. The proton fluxes are the annual sums of integral 5-minute averages for energies $>$10 MeV,  measured by NASA's GOES spacecraft located in Geosynchronous orbit (proton flux source data: NOAA Space Environment Services Center).
}
\end{figure}

A first estimate on general solar activity  can be obtained by looking at the sun spot counts. Displayed in the left panel of Fig.~\ref{fig:solarActivity} is a phase-wrapped plot showing the sunspot counts over the  last six solar cycles. It can be observed that, indeed, the activity of the sun has been low during this cycle, however, not low enough to explain  the discrepancy. However, shown in the right panel of the same figure is a plot showing the annually-accumulated proton fluxes as measured by the GOES spacecraft over the last number of solar cycles. These data are overplotted against the sunspot count values. Here, the element of randomness in the directionality of the proton events is readily apparent, and we can see that, as well as the sun activity being low since launch, we have also been lucky in terms of the number of protons directed towards earth, and therefore L2. We note, however, the possibility that a single large event that is directed towards L2 could completely change the extrapolated values.   See the caption for further detail.

We can, therefore, conclude that the vast majority of the discrepancy between the predicted and observed (so far) CTI levels is due to the low activity of the sun combined with the low number of L2-directed proton events since launch.

\section{CTI DIAGNOSTICS DURING PAYLOAD HEATING EVENTS}
\label{sec:heating_events}

Due to (an ever-decreasing amount of) outgassing and water ice forming in the thermal tent of Gaia, it is necessary to periodically heat the payload and decontaminate the optical surfaces. There have been three such controlled heating events carried out  so far in the mission.  These heating events typically result in an increase in temperature of the FPA of around $25 - 30$~K. During these events we can usually still monitor the charge injection diagnostics during the temperature changes, providing the possibility to learn more about the trap species that have been generated in the devices.

\begin{figure}
	\begin{center}
	\includegraphics[height=6.45cm]{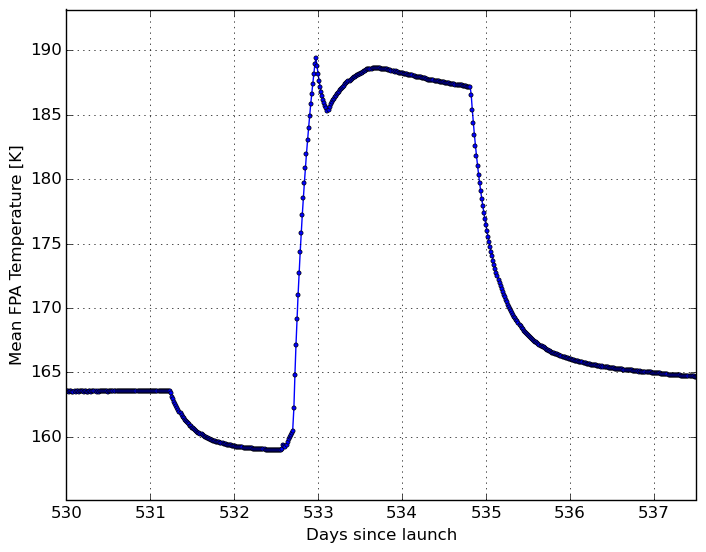}
	\includegraphics[height=6.55cm]{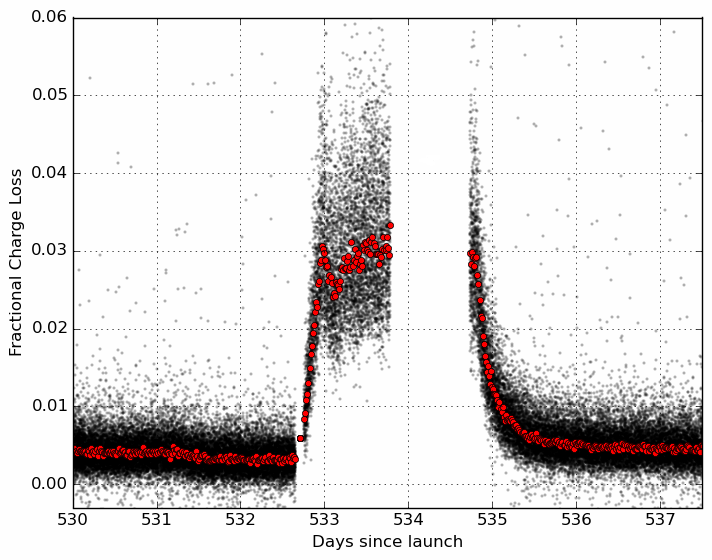}
   \caption{ \label{fig:fprTemperature} 
{\bf{Left:}} Temperature profile (averaged over all sensors close to the FPA) around a controlled heating event performed for telescope decontamination purposes. 
{\bf{Right:}} The response of the fractional charge loss diagnostic for all AF CCDs during the payload heating (red points show a running median for all devices).  It can be observed that the number of trapped electrons increases dramatically with increasing temperature. Note that the data are less noisy than the scatter seems to suggest because the black datapoints are from all AF CCDs, each having different temperatures and levels of radiation damage.
}
	\end{center}
\end{figure}

Displayed in Fig.~\ref{fig:fprTemperature} is the average FPA temperature change during the last decontamination event (left) and the average AF response of the fractional charge loss to the temperature changes (right). It can easily be observed that the temperature increase results in dramatically more trapping from the leading edge of the charge injection profile. 

\begin{figure}
	\begin{center}
	\includegraphics[height=7cm]{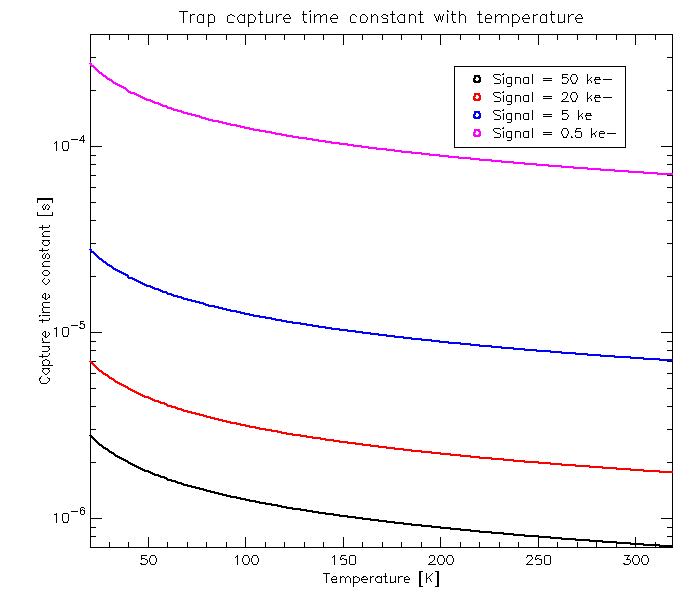}
	\includegraphics[height=7cm]{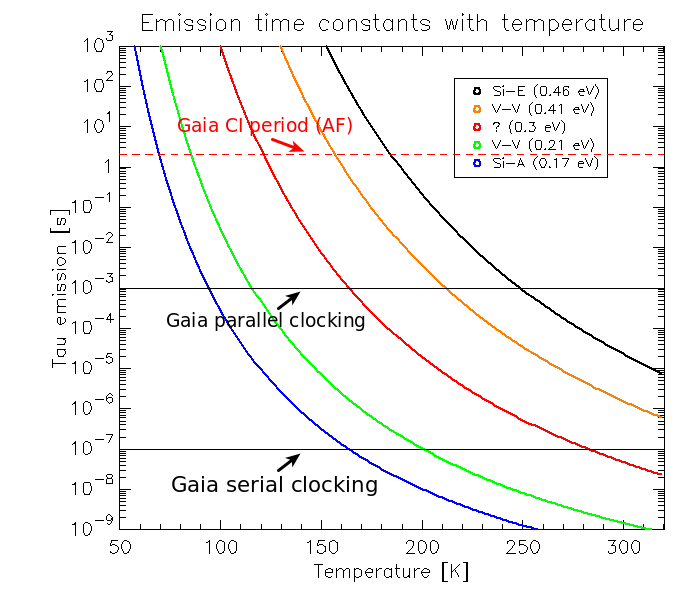}
   \caption{ \label{fig:traps} 
{\bf{Left:}} Theoretical change in the capture time constant for a number of signal levels. It is apparent that this parameter is not very sensitive to temperature changes over the range of the temperature changes during the payload heating events.
{\bf{Right:}} Theoretical change in the characteristic emission time constants for a number of  trap species commonly-referenced in the literature. Also denoted are the parallel and serial clocking periods for the Gaia CCDs, as well as the charge injection period for the AF devices (2~s). Due to the exponential temperature dependence of the emission time constants, traps that  emit on time-scales much longer than the Gaia charge injection period at nominal FPA temperatures will begin to release their trapped electrons within the charge injection period when the temperature changes sufficiently, and therefore become effectively active. This behavior implies that many more traps can be empty at higher temperatures when the next block of charge injection arrives, thus increasing the number of trapped electrons. This behavior can qualitatively explain the large observed increase in fractional charge loss with temperature (Fig.~\ref{fig:fprTemperature}). 
}
	\end{center}
\end{figure}

It is well-known that CTI effects are temperature-dependent. However, the very sharp rise in trapping with temperature may  initially seem surprising due to the fact that, according to Shockley-Read-Hall theory, the capture time constant is only relatively weakly dependent on temperature (see Refs.~\citenum{1974ITED...21..701M} and~\citenum{2001sccd.book.....J} for details on the equations involved for capture and emission). For example, plotted in the left panel of Fig.~\ref{fig:traps} is the expected behavior of the  trapping capture time constant with temperature, for a range of signal levels. However, the temperature effect can be understood in terms of the change in the emission time constant of the traps with temperature where there is an exponential dependency. As the temperature increases, the traps empty much faster. For some traps, this means that they will empty back into the same electron packet from which they were captured, and so will be rendered inactive. For traps with much longer time constants (at nominal operating temperatures) than the 2~s injection period employed onboard the Gaia AF devices, these traps are inactive since they are effectively permanently filled by the injected charge. However, as the temperature is increased, the time constants reduce dramatically, meaning that the traps can release their electrons within the 2~s time-frame, thus making themselves available for trapping at the time of arrival of the following block of injection lines. The temperature dependency of several trap species  commonly referenced  in the literature is plotted in the right panel of Fig.~\ref{fig:traps}. Note that here we model the traps using a single energy level (corresponding to a single emission time constant), however there is evidence that in reality the structure is more like an energy band centered around this energy level, see Ref.~\citenum{2014ITNS...61.1826H} for further detail.

\begin{figure}
	\begin{center}
	\includegraphics[height=6.4cm]{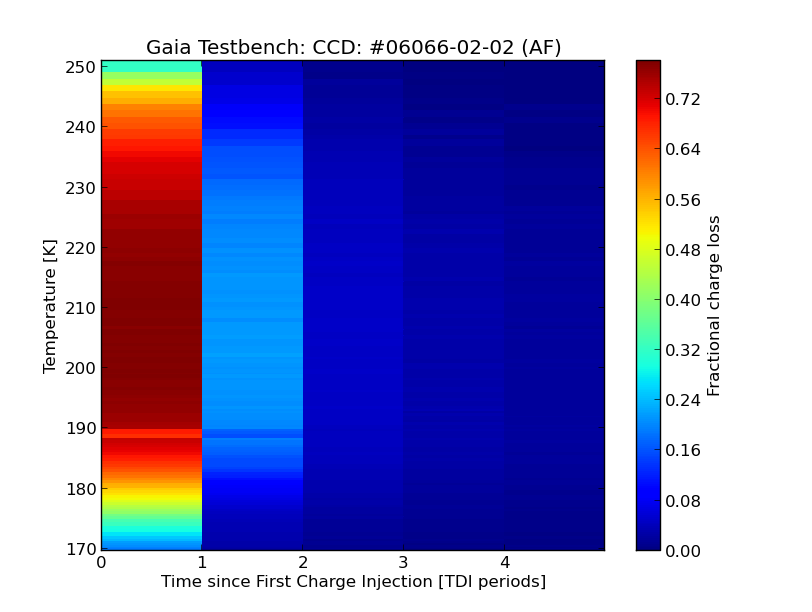}
	\includegraphics[height=6.4cm]{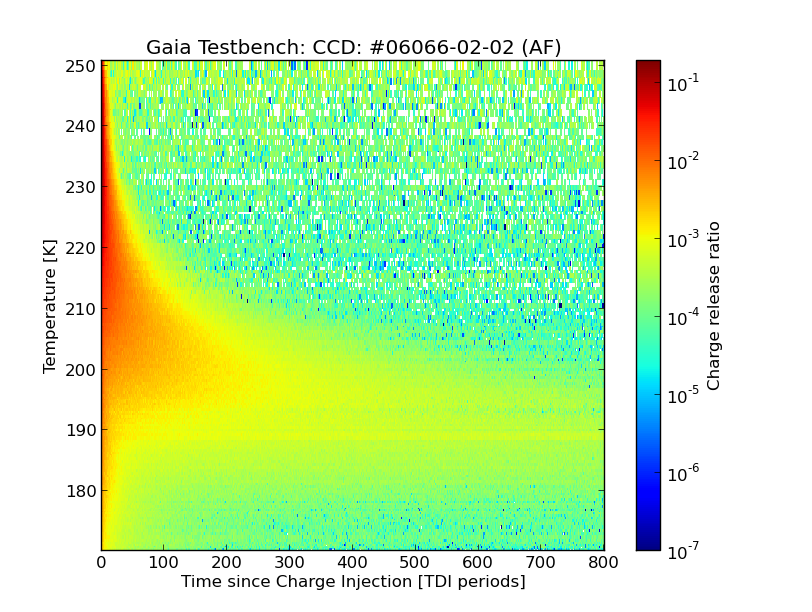}
	\caption{\label{fig:2d} 
{\bf{Left:}} Behavior of the fractional charge loss diagnostic for the first five injected rows as a function of temperature, as measured on-ground using an irradiated engineering model device.
{\bf{Right:}} Profile of the first $800$~rows of the charge release trail as a function of temperature, as measured on-ground.
}
	\end{center}
\end{figure}

In order to explore a wider temperature range, and to compare the temperature dependency in-flight to on-ground (warm-irradiated) data, we measured on-ground the fractional charge loss and de-trapping trail dependencies with temperature using the same irradiated AF device as was discussed earlier in this paper using the Gaia test bench. The results for the trapping (left panel) and trails (right panel) are displayed in Fig.~\ref{fig:2d}.

\subsection{Modelling FPR temperature dependencies}
\label{sec:fpr_heating}

\begin{figure}
	\begin{center}
	\includegraphics[height=6.4cm]{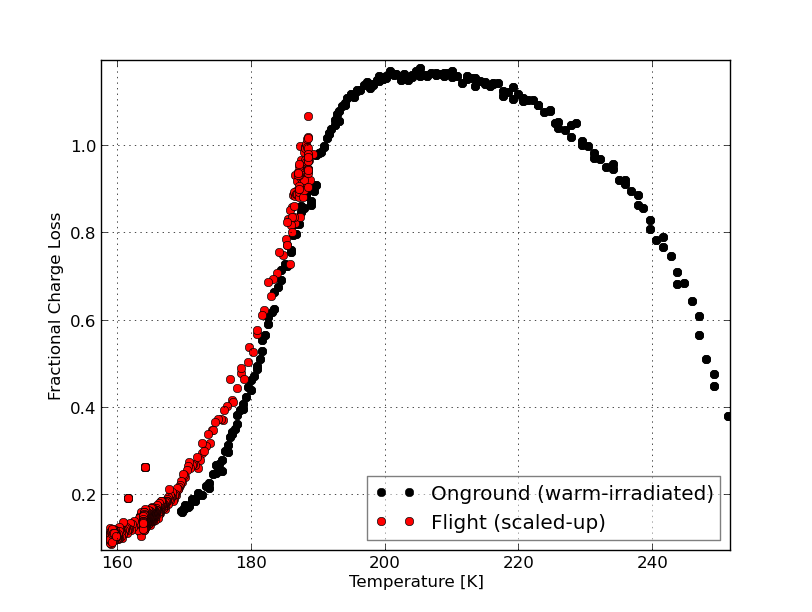}
	\includegraphics[height=6.4cm]{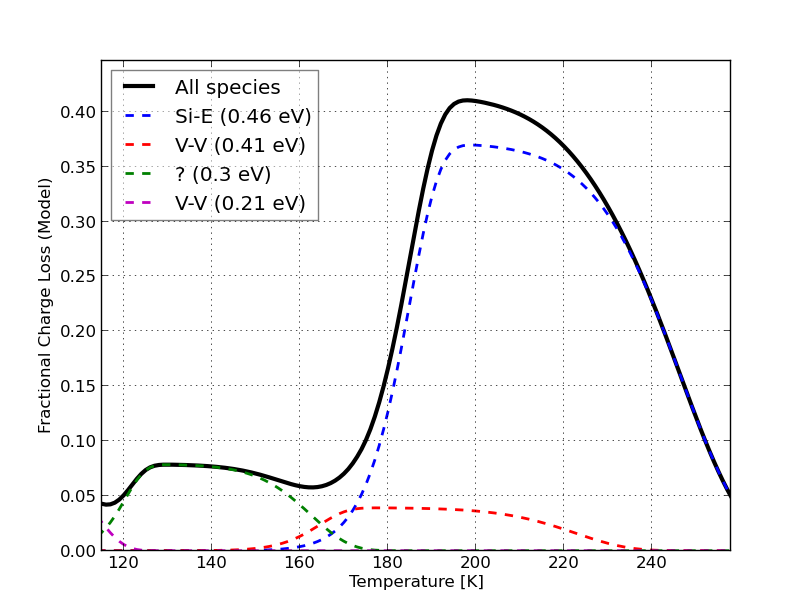}
	\caption{\label{fig:increases} 
{\bf{Left:}} Comparison of the increase in the fractional charge loss with temperature for flight data (black) and data acquired on an on-ground proton-irradiated (at room temperature) engineering model device. It should be noted that the fractional charge loss values rise to greater than unity due to the fact that here we display the sum of the trapped charge over the first number of pixels normalized by the injection level. Also, note that the flight data is scaled-up in order to align the values at the Gaia nominal operating temperature.
{\bf{Right:}} Model predictions of the variation in the fractional charge loss with temperature for Gaia CCDs.  The shape of the profile qualitatively explains the variation observed in the left panel, see text for further detail.
}
\end{center}
\end{figure}

In order to validate the above-described interpretation  of the temperature dependency of the fractional charge loss, the CTI model described in Ref.~\citenum{2013MNRAS.430.3078S} was used to attempt to qualitatively replicate the observed behavior. The same trap species as those described in the previous section were used as input, with their emission times made dependent on the temperature. The results are displayed in Fig.~\ref{fig:increases} (right panel) and can be compared to the on-ground (black) and flight (red) results shown in the left panel. It can be observed that the morphology seen in the data is fairly well matched by the results of the model. Also interesting to note is the impression that there is a slight excess observed in the flight data relative to the on-ground data between $\sim175-180$~K suggesting that the relative concentrations of produced traps that are active at this temperature may be different when irradiated in cold relative to warm. However, more work would be required in order to fully understand this  mis-match quantitatively.

\section{DISCUSSION AND FUTURE WORK}
\label{sec:discussion}

We have shown that the CTI effects (in the parallel direction) observed so far on Gaia are predominantly due to displacement damage caused by GCRs. The lower than expected CTI levels are mostly due to low sun activity combined with the occurrence of very few earth-directed solar flares (two are visible  directly in the CTI diagnostics and it is known, using onboard transient counter data,  that two more took place early in the mission before the charge injection was switched on, see Ref.~\citenum{detector_dr1}). In terms of lessons learned that can be applied to other missions, it is difficult to extrapolate very much regarding the low levels of solar proton damage seen by Gaia, as this is predominantly due to unusually low solar activity. However, it can be noted that the displacement damage caused by GCRs at L2 is not negligible, and, unlike the softer solar protons, the high energy GCRs cannot be effectively shielded against (see Ref.~\citenum{detector_dr1}).

The monitoring of radiation effects (including ionizing radiation effects, displacement damage in the serial registers, hot pixel generation) will of course continue during the course of the Gaia mission. Indeed, as the calibrations improve in the cyclical processing 
 of the science data, the CTI effects on the stellar images themselves will become apparent and need to be taken into account during the image treatment. Additionally, the on-ground test facility will continue to be used to support the interpretation of the flight data and future decontamination events will be taken advantage of to glean any additional information on the trap species that may be available.

\section{CONCLUSIONS}
\label{sec:conclusions}

We have presented a brief overview of the Gaia FPA, its CCDs and their operating conditions at L2. We show that  L2-directed solar activity has been relatively low since launch, and radiation damage (so far) is much less than originally expected. However, an analysis of CTI diagnostics based on the periodic charge injection taking place onboard shows that there are clear cases of correlation between earth-directed solar coronal mass ejection events and abrupt changes in CTI  over time. These sudden jumps are lying on top of a rather constant increase in CTI, which we show is primarily due to the continuous bombardment of the devices by high-energy Galactic Cosmic Rays.  We have also analyzed the changes in the CTI diagnostics as a function of temperature during controlled heating events onboard and show that the findings can be well-modeled by taking into account the temperature dependency of the emission time constants. All $106$ devices are operating well and the low levels of accumulated radiation damage is welcome news regarding the science goals of Gaia.


\acknowledgments 
 
It is a pleasure to thank the Gaia Radiation Taskforce and Payload Expert groups for constructive feedback and advice as  well as the Future missions section at ESTEC for operation of the Gaia Testbench for the detector system. 
We also wish to acknowledge the role of e2v technologies in the detector development process and Airbus Defence \& Space  for the CCD-PEM detector system, the FPA and on-ground testing.

This work has made use of results from the ESA space mission Gaia, the data
from which were processed by the Gaia Data Processing and Analysis Consortium
(DPAC). Funding for the DPAC has been provided by national institutions, in
particular the institutions participating in the Gaia Multilateral Agreement.
The first four authors are members of the Gaia Data Processing and Analysis Consortium
(DPAC). This work has also made use of the SPace ENVironment Information System (SPENVIS), initiated by the Space Environment and Effects Section (TEC-EES) of ESA and developed by the Belgian Institute for Space Aeronomy (BIRA-IASB) under ESA contract through ESA’s General Support Technologies Programme (GSTP), administered by the Belgian Federal Science Policy Office.

\bibliography{report} 
\bibliographystyle{spiebib} 

\end{document}